\documentclass[12pt]{article}
\usepackage{amsfonts}

\makeatletter
\@addtoreset{equation}{section} 
\makeatother

\newcommand{\eqn}[1]{(\ref{#1})}
\def\appendix#1
{
\addtocounter{section}{1}\setcounter{equation}{0}
 \renewcommand{\thesection}{\Alph{section}}
 \section*{Appendix~\thesection\protect\indent \parbox[t]{11.715cm}{#1}}
 \addcontentsline{toc}{section}{Appendix \thesection\ \ \ #1}
}
\newcommand{\real}{{\mathbb{R}}} 

\def\<#1,#2>{\left\langle#1,#2\right\rangle} 

\def\nn{\nonumber}
\newcommand\opname[1]{\mathop{\mathrm{#1}}\nolimits}

\newcommand{\Tr}{\opname{Tr}}

\def\be{\begin{equation}}
\def\ee{\end{equation}}
\def\beqa{\begin{eqnarray}}
\def\eeqa{\end{eqnarray}}
\def\bd{\begin{displaymath}}
\def\ed{\end{displaymath}}

\def\del{\partial}

\newcommand{\bea}{\begin{array}}
\newcommand{\ea}{\end{array}}

\newcommand{\bean}{\begin{eqnarray*}}
\newcommand{\eean}{\end{eqnarray*}}

\begin{document}
\begin{titlepage}
\begin{flushright}
\baselineskip=12pt
DSFNA-16/04\\
\hfill{ }\\

\end{flushright}

\begin{center}

\baselineskip=24pt

{\Large \bf Duality symmetry for star products}
\baselineskip=14pt

\vspace{1cm}

{\bf V. I. Man'ko} $^{a,b}$, {\bf G. Marmo} $^{b}$,
 and {\bf P. Vitale} $^{b}$\\[6mm]
$^a$ {\it
P. N. Lebedev Physical Institute ,
Leninsky Pr. 53, Moscow, Russia
}\\[6mm]
$^b$ {\it Dipartimento di Scienze Fisiche, Universit\`{a} di Napoli {\sl
Federico II}\\
and {\it INFN, Sezione di Napoli, Monte S.~Angelo}\\
Via Cintia, 80126 Napoli, Italy}\\
{\tt manko@na.infn.it, giuseppe.marmo@na.infn.it,patrizia.vitale@na.infn.it}
\\
\end{center}

\vskip 2 cm

\begin{abstract}
A  duality property for star products is exhibited. In view of it,
known star-product schemes, like the Weyl-Wigner-Moyal formalism,
the Husimi and the Glauber-Sudarshan maps are revisited and their
dual partners elucidated. The tomographic map, which has been
recently described as yet another star product scheme, is
considered. It yields a noncommutative algebra of operator symbols
which are positive definite probability distributions. Through the
duality symmetry a new noncommutative algebra of operator symbols
is found, equipped with a new star product. The kernel of the new
star product is established  in explicit form and examples are
considered.
\end{abstract}
\end{titlepage}

\section*{Introduction}
Star-products, originally introduced in the context of the
quantization  problem, have recently attracted new interest partly
because of their appearance in Noncommutative Geometry.
  Among the different star-product schemes known in the literature
  since many years,  the
Weyl-Wigner-Moyal formalism \cite{We31,Wi32,Gr46}, which consists
of an invertible map from
 (Hilbert-Schmidt) operators onto (Schwartzian) functions on
$\real^{2n}$ with the Moyal product, plays a distinguished role.
Various modifications of such type of maps exist yielding new
operator symbols, analogous to Wigner functions,  known in the
literature as the Husimi Q-function \cite{Hu40}, the P-function
introduced in \cite{Su63, Gl63} and s-ordered quasi-probability
distributions \cite{CG69}. In a modern language we recognize these
sets of symbols as noncommutative associative algebras equipped
with a noncommutative (star) product.

As anticipated above, the interest in star products, which have
been intensively studied for long time, has recently received new
impulse (a very partial list of references  is
\cite{Berezin,BFFLS78,VGB89,GB92,DMO92,Fe96,Ko97,CUZ01,KO02,Da03}).
One of the reasons
 is certainly the emergence of  Noncommutative
Geometry  in relevant sectors of physics like field theory, where
it came out that the dynamics of  strings in the presence of a
magnetic field is described by a noncommutative geometry
associated to the Moyal product \cite{strings}. Since then, field
theories on  Moyal-type spaces have been scrutinized extensively;
good reviews are~\cite{DouglasN,Szabo}. The new products are most
often either products of finite matrices (the fuzzy algebras
\cite{fuzzy}) or can be reduced to the Moyal product. Recently,
 \cite{BLMV01}  a general method  is proposed, which
produces various new non-formal star products on~$\real^3$ using a
variation of the Jordan Schwinger map \cite{MMVZ94}. The method
includes previous results of \cite {HLS01} and it is easily
extendable to higher dimensions \cite{ALZ02}.

In a different setting it was recently established
\cite{Marmo1,Marmo2} that the symplectic \cite{MMT95,MMT96} and
spin  \cite{DM97, MM97} tomography, which furnish alternative
formulations of quantum mechanics and quantum field theory
\cite{MRV98}, can be described as well within  a star-product
scheme. Moreover, in \cite{Marmo2} different known star-product
schemes were presented in a unified form. There, the symbols of
the operators are defined in terms of a special family of
operators using the trace formula (what we sometimes call the
`dequantization' map because of its original meaning in the
Wigner-Weyl formalism), while the reconstruction of operators in
terms of their symbols (the `quantization' map) is determined
using another family of operators. These two families determine
completely the star-product scheme, including the kernel of the
star-product.

The two families of operators allow to express any of the
operators belonging to  the  space of linear operators as a
combination of them. Moreover,  they are dual to each other.
Indeed,  the key observation of the present article is that their
role can be exchanged, without violating the consistency of the
scheme. This implies that the
 family of operators originally used in the reconstruction
map (from functions to operators) can be used instead in the
`dequantization' map to determine a different set of symbols which
is a different noncommutative associative algebra, dual to the
original one, with a new star product. This duality symmetry is
not specific to the tomographic map. The latter  is just an
example where it produces a new noncommutative algebra, whose
nature is still to be fully understood. It is instead a property
of the star-product scheme as it appears in the unifying form
described in \cite{Marmo2}. The symmetry could yield to
interesting results in other relevant cases as those considered in
\cite{BLMV01,ALZ02}, which we plan to investigate later.

The paper is organized as follows. In sec. 1 we review the general
building scheme for the  star-product of operator symbols using
two families of basic operators. In sec. 2 we introduce the
duality symmetry and the dual star product. In section 3 we
investigate the duality property of known star-product  schemes.
The Weyl-Wigner-Moyal formalism will result to be self-dual, as
already known in  various forms, while the so called s-ordered
symbols will show a  duality relation between normal-ordered (s=1)
and antinormal-ordered (s=-1)  symbols. In section 4 we review the
tomographic map (symplectic tomography) and the star-product
construction of the operator-tomograms. In Sec. 5 we introduce the
dual tomographic map and study its properties. Moreover, we derive
the kernel of the star-product for the dual tomographic symbols.
Although the noncommutative algebra of symbols and the related
star-product where already established, the dual algebra and the
new star-product, discovered thanks to the duality symmetry, were
not known to our knowledge. Conclusions and perspectives are given
in Sec. 6.

\section{The star product}
In this section we review  a
 general construction  for  the symbols of
operators acting in a Hilbert space $\mathcal H$. We follow the
presentation and the notation of \cite{Marmo2}.

Given an operator $\hat A$ acting on the Hilbert space $\mathcal
H$ (which can be finite or infinite-dimensional), let us have two
basic families $\hat U(\bar x)$ and $\hat D (\bar x)$ acting in
$\mathcal H$. The families are labelled by a vector $\bar x=(x_1,
...x_N)$. The parameters $x_k, (k=1,...,N)$ can be continuous
(real or complex) or discrete. The symbol $f_A(\bar x) $ of the
operator $\hat A$ is the function depending on the vector $\bar x$
defined by the formula \be f_A(\bar x)= \Tr (\hat A\hat U(\bar
x)). \label{deq-map} \ee We assume that the trace exists for all
the parameters $\bar x$. The introduced function is understood as
a generalized function.  The second family of operators, $\hat
D(\bar x)$, serves to reconstruct the operator if one knows its
symbol, that is  \be \hat A=\int f_A(\bar x) \hat D(\bar x) d\bar
x . \label{quant-map} \ee We assume that an appropriate measure
$d\bar x$ exists to make sense of  the reconstruction formula. If
there are discrete components $x_k$ the integral \eqn{quant-map}
splits into a sum over the discrete components and an integration
over the remaining continuous ones.

 The symbols form an associative algebra endowed with a
non-commutative (star) product inherited by the operator product.
The associativity follows from the associativity of the operator
product $\hat A(\hat B\hat C)=(\hat A\hat B)\hat C$. This defines
for the symbols the star product \be f_A\ast f_B (\bar x)= f_{AB}
(\bar x)=  \int f_A(\bar x_1) f_B (\bar x_2) K(\bar x_1,\bar x_2,
\bar x) d\bar x_1 d\bar x_2, \label{stprod} \ee where the kernel
$K$ must satisfy the nonlinear equation \be  \int  K(\bar x_1,
\bar y, \bar x)  K(\bar x_2,\bar x_3, \bar y) d\bar y = \int
K(\bar x_1,\bar x_2, \bar y)  K(\bar y,\bar x_3, \bar x) d\bar y
\label{intcond} \ee obtained by imposing the associativity
condition.
 Let us get the
expression of the kernel of the star product in terms of the
operators $\hat U(\bar x)$ and $\hat D(\bar x)$. To obtain it we
take \eqn{quant-map} for an operator $\hat A$ and multiply it on
the right by an analogous expression for an operator $\hat B$.
Then we calculate the symbol of this product by means of
\eqn{deq-map} and we obtain \be f_{AB} (\bar y)=\int \Tr
\left(f_A(\bar x) \hat D(\bar x)f_B (\bar x') \hat D(\bar x') \hat
U(\bar y)\right) d\bar x d\bar x'. \ee Comparing this expression
with \eqn{stprod} we have \be K(\bar x_1,\bar x_2, \bar y)=
\Tr(\hat D(\bar x_1) \hat D(\bar x_2) \hat U(\bar y)).
\label{kernel} \ee Because of the construction of the kernel and
the associativity of the operator product we are guaranteed that
\eqn{kernel} satisfies the nonlinear equation \eqn{intcond}.

Writing the formula \eqn{deq-map} in view of \eqn{quant-map} one
has \be f_A(\bar x)= \Tr\left\{\left[\int f_A(\bar x') \hat D(\bar
x') d\bar
  x'\right]  \hat U(\bar x)\right\}
\ee
and, assuming we can exchange the trace with the integral,
 we arrive at the equality
\be
f_A(\bar x)= \int d\bar x'\left[ \Tr \hat D(\bar x') \hat U(\bar
  x)\right]  f_A(\bar x'). \label{deq2-map}
\ee  The compatibility condition for Eqs.
\eqn{deq-map}-\eqn{deq2-map} requires then
  \be
\Tr \hat D(\bar x') \hat U(\bar  x) =  \delta(\bar x'-\bar x),
\label{compatibility} \ee where $ \delta$ is to be replaced with
the  Kronecker delta for discrete parameters. Going back to
\eqn{kernel}, it is to be stressed that the kernel of the star
product has been obtained solely in terms of the operators $D(\bar
x)$ and $U(\bar x)$, which in turn are only constrained by
\eqn{compatibility}. Thus, to each pair of operators satisfying
\eqn{compatibility} it is associated a associative algebra with a
star product. The role of $D(\bar x)$ and $U(\bar x)$ can be
exchanged, what gives rise to a duality symmetry.
\section{The dual map and the dual star-product}
  The two
families of operators $\hat U(\bar x)$ and   $\hat D\bar x)$ are
connected by the equation \eqn{compatibility}. This implies that
they can be interpreted as orthogonal families  in the space of
operators. In facts, it is well known that the linear structure in
the space of operators (matrices) allows to consider them as
vectors (see for example \cite{MMSZ03}). The trace of two
operators is equivalent then to the standard scalar product in the
linear vector space. In this context the symbol of an operator
defined by means of the family $\hat U$ can be treated as a
projection onto the  `vector' $\hat U_{\bar x}\equiv \hat U(\bar
x)$. But the `vectors'$\hat D_{\bar x}\equiv \hat D(\bar x)$
provide a dual family since their scalar product with the vectors
$\hat U_{\bar x}$ is equal to a delta function. In view of this
geometrical picture it is obvious that one can define a dual
symbol $f_A^{(d)}(\bar x)$ of an operator $\hat A$ using the
projection on the dual family $\hat D_{\bar x}$, i.e. \be
f_A^{(d)}(\bar x)=\Tr(\hat A \hat D (\bar x)). \label{dualdeq-map}
\ee The reconstruction formula for the operator $\hat A$ reads
then \be \hat A=\int f_A^{(d)}(\bar x) \hat U (\bar x) d\bar x
\label{dualquant-map} \ee which means that we have exchanged the
role of $\hat U$ and $\hat D$ in our previous formulae
\eqn{deq-map} and \eqn{quant-map}. This duality property being
simple,  provides new nontrivial symbols and a new associated star
product with a new kernel in all cases where the operators $\hat
U$ and $\hat D$  are essentially different. From Eq. \eqn{kernel}
the expression for the kernel of the dual star product is easily
derived to be \be K^{(d)}(\bar x_1,\bar x_2, \bar y)\equiv
\Tr\left(\hat U(\bar x_1) \hat U(\bar x_2) \hat D(\bar y)\right).
\label{dualkernel} \ee

\section{The Weyl-Wigner and the  s-ordered maps}
In this section we review known examples  of
quantization-dequantization maps from the viewpoint of the
duality property illustrated above.

Given an operator $\hat A$ acting in the Hilbert space of square
integrable functions $\psi(x)$ on $\real$, the Wigner-Weyl symbol
of $\hat A$ and the reconstruction map are defined by means of the
families of operators \beqa \hat U(\bar x)&=&2\hat {\mathcal
D}(\alpha) (- \hat 1)^{a^\dag a} {\mathcal
  D}(-\alpha) \label{wigner}\\
\hat D(\bar x)&=&{1\over 2\pi} \hat U(\bar x) \label{weyl} \eeqa
where $(-\hat 1)^{a^\dag a}$ is the parity operator. We may use a
symplectic notation for the real vector and write $\bar x=(q,p)$.
The annihilation and creation operators $a$ and $a^\dag$ read \be
a= {1\over \sqrt 2} (\hat q+ i\hat p),~~~~ a^\dag = {1\over \sqrt
2} (\hat q-i\hat p). \label{aadag} \ee The unitary displacement
operator (complex Weyl system) realizing the ray representation of
the group of translations of the plane is of the form \be
\hat{\mathcal D} (\alpha) = \exp(\alpha a^\dag-\alpha^\ast a)
\label{displacement} \ee with $\alpha=(q+i p)/\sqrt{2}$. Using
known properties of this operator \beqa (-\hat 1)^{a^\dag a} \hat
{\mathcal D}(\alpha) (-\hat 1)^{a^\dag a}&=& \hat {\mathcal
D}(-\alpha) \label{reflex}\\
 \Tr\hat {\mathcal D}(\alpha)&=&\pi
\delta({\rm Re}~\alpha) \delta({\rm Im}~ \alpha)
\label{displtrace}\eeqa one can check that \be \Tr  \hat U_{\bar
x} \hat D_{\bar x'}=\delta(q-q')\delta(p-p'). \ee The Weyl symbol
of the operator $\hat A$ reads \be f_A(q,p) =\Tr(\hat U(\bar x)
\hat A) \ee which in coordinates representation becomes \be
f_A(q,p) =\int A(q+{u\over 2}, p-{u\over 2} )e^{-ipu} du \ee with
$A(x,x')=\langle x|\hat A|x'\rangle$. For the density operator
$\hat \rho$ of a normalized state the Weyl symbol is exactly what
is known as the Wigner function of the quantum state \be W(q,p)
=\int \rho(q+{u\over 2}, p-{u\over 2} )e^{-ipu} du . \ee Moreover
the Weyl symbol of the identity operator is $f_1(q,p)=1$ where $1$
is the identity in the algebra of functions on $\real^2$, whereas
the Weyl symbols of the position and momentum operators are
$f_q(q,p)=q$, $f_p(q,p)=p$.  The operator $\hat A$ can be
expressed in terms of its Weyl symbol by means of the
reconstruction formula \eqn{quant-map} \be \hat A= {1\over \pi}
\int f_A(q,p)\hat {\mathcal D}(\alpha)(-\hat 1)^{a^\dag a}
     \hat {\mathcal D}(- \alpha) dq dp \label{quanweyl}
\ee which, specialized to the density operator, yields the well
known relation between the density operator and the Wigner
function. In coordinates representation it reads \be
\rho(x,x')={1\over 2\pi}\int W({x+x'\over 2}, p)e^{ip(x-x')} dp.
\ee The algebra of symbols is what is known as the Moyal plane,
that is a noncommutative algebra of functions on $\real^2$ with
the Moyal product. The kernel is easily obtained specializing
\eqn{kernel} to this case and using Eq. \eqn{reflex} together with
the product rule for the displacement operators
 \be \hat {\mathcal D}
 (\alpha)\hat {\mathcal D}(\beta)={ \mathcal D}(\alpha+\beta)
e^{i~Im(\alpha\beta^\ast)} \ee Thus we reduce the calculation of
the kernel to evaluate a trace of the form \eqn{displtrace}. The
final result may be written in the form
 \be K(q_1,p_1,q_2,p_2,q_3,p_3)= {1\over \pi^2} \exp\left\{2i\left[q_1(p_2-p_3)
 + q_2(p_3-p_1)+ q_3(p_1-p_2)\right]\right\} \label{moykernel1}
 \ee
or equivalently
\be
 K(\alpha_1,\alpha_2,\alpha_3)= {1\over \pi^2}\exp\left\{4i
{\rm Im} \left[ (\alpha_1^\ast\alpha_2 + \alpha_3\alpha_2^\ast
+\alpha_1\alpha_3^\ast) \right] \right \}. \label{moykernel}\ee
  As from \eqn{wigner},
  \eqn{weyl},  in the
Wigner-Weyl-Moyal formalism the operators $\hat U(\bar x) $ and
$\hat D(\bar x)$ coincide up to a numerical factor, which implies
that this scheme is selfdual. That is,  the duality operation
doesn't produce a new algebra of symbols nor  a new star-product.
\subsection{s-ordered symbols}
In \cite{CG69}  s-ordered quasidistribution functions
 which are s-ordered symbols of density operators were introduced as a
 generalization of the Wigner, Husimi and Glauber-Sudarshan distribution functions.
Later on it has been realized that these symbols, together with
their reconstruction formulas,  may be
 obtained through the two families of operators $\hat U_s(x)$ and
 $\hat D_s(x)$, $\bar x=(q,p)$ of the form
\beqa \hat U_s(x)&=&{2\over (1-s)} \hat {\mathcal D}(\alpha)
\left({s+1\over
 s-1}\right)^{a^\dag a} \hat {\mathcal D}(-\alpha)\\
\hat D_s(x)&=&{1\over 2\pi} \hat U_{-s}(x) \label{sordop} \eeqa
where $s$ is a real parameter. Moreover s-ordered symbols may be
considered not only for the density operator but for a generic
operator $\hat A$ according to the general scheme presented in the
previous sections. The case $s=0$ corresponds to the standard
$Wigner-Weyl$ situation described above. In
 the limit $s=\pm 1$ we have respectively the diagonal representation of
 the density matrix (P-function) by Sudarshan \cite{Su63} and Glauber
 \cite{Gl63}  and the Q-quasidistribution \cite{Hu40}.
  The star-product kernel of s-ordered symbols is calculated along
  the same lines than  the Moyal kernel \eqn{moykernel}, thus yielding~\cite{Marmo2}
\beqa
 K(\alpha_1,\alpha_2,\alpha_3)&=& {1\over (1-s^2)
 \pi^2} \exp\left\{ {2\over 1+s}
 (\alpha_1^\ast\alpha_2 - \alpha_2\alpha_3^\ast
-\alpha_3\alpha_1^\ast) \right. \nn\\&~& \left. + {2\over 1-s}
 (  \alpha_3\alpha_2^\ast
+\alpha_1\alpha_3^\ast -\alpha_2^\ast\alpha_1)+ {4s\over s^2-1}
|\alpha^3|^2 \right \} \eeqa to be compared with \eqn{moykernel}.
It is important to stress that  the two limiting cases
corresponding to $s=\pm 1$ are in the duality relation discussed
 previously. In particular, the Husimi Q-function and the P-function
 are dual symbols of the density operator in such limit cases. They
 belong to two different associative noncommutative algebras, with
 different star-products. It is already known that the two
 quantum schemes correspond to normal and antinormal ordering of
 the creation and annihilation operators. The duality symmetry then
 connects the two orderings. The Wigner-Weyl Moyal quantization scheme
 selects instead the symmetric ordering, consistently with its being
 selfdual.

Another simple example of  a consistent pair of operators $\hat
U(\bar x), \hat D(\bar x)$ producing a noncommutative algebra and
a star product   in the
 sense of \eqn{deq-map}, \eqn{stprod},  is provided by the pair
\be \hat U(\bar x) ={1\over \sqrt{2\pi}} e^{i\mu \hat q+i\nu \hat
p},~~~~ \hat D(\bar x)=  \hat U^\dag (\bar x) \label{fourier} \ee
with $\bar x=(\mu,\nu)$, $\mu,\nu,$ real. This construction
amounts to consider the Fourier transform of the
 Weyl symbol of the operator $\hat A$. That is, the symbol
 $f_A(\mu,\nu)=\Tr(\hat U(\bar x) \hat A)$ is related to the Weyl
 symbol as
\be
f_A(\mu,\nu)={1\over\sqrt{2\pi} } \int f_A(q,p) e^{i\mu q+i\nu p} dq
 dp \label{Wefourier}.
\ee The reconstruction formula reads instead \be \hat A =  {1\over
\sqrt{2\pi}} \int e^{-i\mu \hat q -i\nu \hat p} f_A(\mu,\nu) d\mu
~ d\nu. \label{reconstrfourier} \ee It is evident from Eq.
\eqn{fourier} that the dual star-product construction corresponds
to the replacement
 $\mu,\nu \rightarrow -\mu,-\nu$ or, from Eq. \eqn{Wefourier},
   to the replacement $f_A(q,p) \rightarrow f_A(-q,-p)$. The
 kernel of the star-product for the introduced symbols can be
 calculated as the trace of the product of three exponents
\be K(\mu_1,\nu_1,\mu_2,\nu_2,\mu_3,\nu_3) = {1\over
 \left(\sqrt{2\pi}\right)^3} \Tr (e^{i\mu_3\hat q+i\nu_3\hat p}
e^{-i\mu_1\hat q-i\nu_1\hat p } e^{-i\mu_2\hat q-i\nu_2\hat p })
\ee which yields \beqa &&K(\mu_1,\nu_1,\mu_2,\nu_2,\mu_3,\nu_3)=
(2\pi)^{3/2} \delta(\mu_1+\mu_2-\mu_3)
\delta(\nu_1+\nu_2-\nu_3)\nn\\
&& \exp\left[{i\over 2}\left(\mu_3\nu_1-\mu_1\nu_3 +
\mu_3\nu_2-\mu_2\nu_3 +
 \mu_2\nu_1 -\mu_1\nu_2\right)\right]. \label{reconstform}
\eeqa
\section{The tomographic map}
 The tomographic map was
originally introduced in \cite{MMT95,MMT96} to solve an old issue
in quantum mechanics, namely to  provide a description of quantum
mechanics which were directly in terms of probability
distributions (see also \cite{MRV98,MRV98a}).  Because of its
original scope the map with its inverse was in the beginning
proposed only for the density operators,  as it was the case for
the distribution functions of Wigner, Glauber-Sudarshan and
Husimi.\footnote{In such context it is also referred as the {\it
probability representation} of quantum mechanics because the
quantum states are described by  families of probabilities.}
 It was immediately clear that the space of distributional
 functions associated to density operators
(the density matrix tomograms) is not a commutative one but it
took some time to realize that it could be described within a
star-product scheme \cite{Marmo1, Marmo2}. In \cite{Marmo1,
Marmo2} it was also understood that the scheme is actually more
general, that is, not only density operators but a generic
operator  can be mapped into tomograms with a well defined
associative star-product, thus exhibiting a new example of
noncommutative algebra which might be interesting {\it per se},
independently from its original connection to quantum mechanics.

For the tomographic map the two families of operators $\hat U(x)$
and $\hat D(x)$ have been shown to be
 \beqa
\hat U(\bar x) &=&\delta(X\hat 1-\mu \hat q-\nu \hat p) \label{tomU}\\
\hat D(\bar x) &=&{1\over 2\pi} \exp\left[i(X\hat 1-\mu\hat q -\nu
\hat p)\right] \label{tomD} \eeqa where  ${\bar x}= (X,\mu,\nu)
\in \real^3$. Thus one can prove \cite{MMV01} that the tomographic
symbol of the density operator, originally introduced in
\cite{MMT95,MMT96} in the form
 \be {\mathcal W}_{\rho}(X,\mu,\nu,t)= {1\over
2\pi} \int dk\, e^{-ik X} \Tr (\hat\rho e^{ik\hat X})~ ,
\label{mdf} \ee with $X=(\mu q + \nu p),~  \hat X=(\mu \hat q +
\nu \hat p)$,
 may be reexpressed by means of Eqs. \eqn{deq-map}, \eqn{tomU}   as
 \be {\mathcal W}_\rho(X,\mu,\nu)= \Tr \left( \hat\rho\delta(X\hat 1-\mu
\hat q-\nu \hat p) \right)\label{wrho} \ee where we are adopting
the notation ${\mathcal W}_A$ for the tomographic symbols.  From
this formula it may be directly checked that ${\mathcal W}_\rho$
is a probability density, that is
$${\mathcal W}_\rho \ge 0, ~~{\rm and} \int {\mathcal
W}_\rho(X,\mu,\nu) dx=1.$$ According to Eq. \eqn{stprod} the star
product of two tomographic symbols ${\mathcal W}_A$,${\mathcal
W}_{B}$ is \beqa &&{\mathcal W}_A \ast {\mathcal W}_B (X,\mu,\nu)=
 \int dX_1~ d\mu_1 ~d\nu_1~ dX_2~ d\mu_2 ~d\nu_2 \nn\\
&& ~~~~{\mathcal W}_A(X_1,\mu_1,\nu_1) {\mathcal W}_B
 (X_2,\mu_2,\nu_2) K(X_1,\mu_1,\nu_1,X_2,\mu_2,\nu_2, X,\mu,\nu)
\eeqa where the kernel $K $, defined in Eq. \eqn{kernel}  is given
by
\beqa
 &&K(X_1,\mu_1,\nu_1,X_2,\mu_2,\nu_2,
X,\mu,\nu)=\Tr\left(\hat D(X_1,\mu_1,\nu_1) \hat
D(X_2,\mu_2,\nu_2 ) \hat U(X,\mu,\nu)\right)\nn\\
&&~~~= {1\over 4\pi^2} \exp\left\{{i\over
2}\left[(\nu_1\mu_2-\mu_1\nu_2 +2X_1 +2 X_2)
-\left({\nu_1+\nu_2\over \nu} +{\mu_1+\mu_2\over\mu} \right)X
\right]\right\}\nn\\
&& ~~~~~~~\delta\left(\mu(\nu_1+\nu_2)-\nu(\mu_1+\mu_2)\right).
\label{kerneltom}\eeqa In order to verify the compatibility
condition \eqn{compatibility} we can calculate the trace of the
product of the operators $\hat U(\bar x)$ and $\hat D(\bar x)$. It
reads \beqa \Tr[\hat U(\bar x)\hat D(\bar x')]&=&{1\over 2\pi}
e^{iX'}\Tr \left[e^{-i\mu'\hat q -i\nu' \hat p}
 \delta(X\hat 1- \mu\hat q -\nu \hat p)\right]\nn\\
 &=&{1\over 2\pi}\delta(\nu\mu'-\mu\nu')\exp\left\{{i\over
 2}\left[
 2X'-\left({\mu'\over\mu}+{\nu'\over\nu}\right)X\right]\right\}
 \eeqa
 Although expressed in unusual form, this is a three dimensional delta function
 in the space of functions of $(X,\mu,\nu)$. It can be easily verified observing that
 \be
 \int f(X',\mu',\nu') {1\over 2\pi}\delta(\nu\mu'-\mu\nu')\exp\left\{{i\over
 2}\left[ 2X'-\left({\mu'\over\mu}+{\nu'\over\nu}\right)X\right]\right\} dX'~d\mu' ~ d\nu'
= f(X,\mu,\nu). \ee

\section{Dual tomographic map}
In this section we address the problem of constructing the dual
tomographic map by means of the duality symmetry previously
exploited. Following the prescription, the new families of basic
operators $\hat U$ and $\hat D$ are obtained exchanging the role
of \eqn{tomU} and \eqn{tomD}, that is, we have now \beqa \hat
U_d(\bar x) &=&{1\over 2\pi} \exp\left[i\left(X\hat 1-\mu\hat q
-\nu \hat p\right)\right]
\label{dualtomU}\\
\hat D_d(\bar x) &=& \delta(X\hat 1-\mu \hat q-\nu \hat
p)\label{dualtomD}. \eeqa This means that we associate with the
operator $\hat A$ the symbol \be {\mathcal W}_A^{(d)} (X,\mu,\nu)
= {1\over 2\pi} e^{iX} \Tr (\hat A e^{-i\mu\hat q-i\nu\hat
p}).\label{dualsymbol} \ee One can see that this function differs
from the symbol determined by Eq. \eqn{Wefourier} just by the
factor $[\exp (iX)]/\sqrt{2\pi} $. The reconstruction formula for
the dual tomographic map reads \beqa
 \hat A&=& {1\over 2\pi} \int e^{iX}
\left (\Tr\hat A e^{-i\mu\hat q -i\nu \hat p}\right) \delta(X\hat
1- \mu\hat q -\nu \hat p) dX~d\mu~d\nu\nn\\ &=& \int {\mathcal
W}_A^{(d)} (X,\mu,\nu) \delta(X\hat 1- \mu\hat q -\nu \hat p)
dX~d\mu~d\nu. \label{d4} \eeqa Integrating over dX we obtain \be
\hat A= {1\over 2\pi}\int \left [\Tr\hat A e^{-i\mu\hat q -i\nu
\hat p}\right] e^{i\mu\hat q +i\nu \hat p}
 ~d\mu~d\nu
\ee which is exactly  Eq. \eqn{reconstrfourier}.

The
 dual tomographic symbols satisfy the constraint \be
{\del \over \del X} \left(e^{-iX} {\mathcal W}_A^{(d)}
(X,\mu,\nu)\right) =0 \ee namely the dependence on $X$ is
factorized and it is just a phase.

 Now we can calculate the
kernel of the
 star product as
 \be
 {K}^{(d)}(\bar x_1,\bar x_2,\bar x_3)={1\over 2\pi}e^{iX_3} \Tr\left[
 \delta(X_1\hat 1- \mu_1\hat q -\nu_1 \hat p) \delta(X_2\hat 1-
 \mu_2\hat q -\nu_2 \hat p)\delta(X_3\hat 1- \mu_3\hat q -\nu_3 \hat p) \right]
\ee Using the Fourier decomposition for the delta function we
arrive at
\be
 {K}^{(d)}(\bar x_1,\bar x_2,\bar x_3)={1\over 4\pi^2|S_{12}|}
 \exp\left({i\over S_{12}} V -{i\over 2 }{S_{23}S_{31}\over S_{12}}\right)
 \ee
 where the volume V and the symplectic areas $S_{ik}$ read
 respectively
 \beqa
 V&=&\bar X[\bar\mu{\times} \bar\nu]\\
 S_{ij}&=&\mu_i\nu_j-\mu_j\nu_i
 \eeqa
 and $\bar X=(X_1,X_2,X_3),~
 \mu=(\mu_1,\mu_2,\mu_3),~\nu=(\nu_1,\nu_2,\nu_3)$.
 The kernel for the $\ast$ commutator of dual tomographic symbols is then
\beqa
 {\mathcal K}^{(d)}(\bar x_1,\bar x_2,\bar x_3)&=&{K}^{(d)}(\bar x_1,
 \bar x_2,\bar x_3)-{K}^{(d)}(\bar x_2,\bar x_1,\bar x_3)\nn\\
 &=&
 {i\over 2\pi^2 |S_{12}|}
 \exp\left({i V\over S_{12}}\right) \sin\left({S_{23}S_{31}
 \over 2 S_{12}}\right)
 \eeqa
which in turn defines a Poisson bracket on the algebra of dual
tomograms, if the commutative limit is performed. We will deepen
this analysis  elsewhere.

\subsection{Examples}
In order to better understand what is the dual tomographic map it
is worth establishing the form  of dual tomographic symbols for
some relevant operators. From Eq. \eqn{dualsymbol}  we have for
the  identity operator \be {\mathcal W}_1^{(d)}(X,\mu,\nu)=e^{iX}
\delta(\mu) \delta(\nu). \ee This is the identity in the space of
dual tomograms.  Analogously we find for  powers of momentum and
position operators \beqa
{\mathcal W}_{q^n}^{(d)}(X,\mu,\nu)&=&(i)^n e^{iX}\,\delta^{(n)}(\mu) \delta(\nu)\\
{\mathcal W}_{p^n}^{(d)}(X,\mu,\nu)&=&(i)^n e^{iX}\,\delta(\mu)
\delta^{(n)}(\nu) \eeqa where $\delta^{(n)}$ is the $n-th$
derivative of the Dirac delta function.

The density operator $|\alpha\rangle\langle\alpha|$ of  the
coherent state $|\alpha\rangle$, is mapped to \be {\mathcal
W}_\alpha^{(d)}(X,\mu,\nu)={e^{iX}\over 2\pi}
e^{-{(\mu^2+\nu^2)\over 4}} \exp\left[ i{\rm
Im}\left(\alpha^*{(\nu-i\mu)\over \sqrt{2}}\right)\right] \ee
while the density operator  $|n\rangle\langle n|$ of the Fock
state $|n\rangle$ corresponds to \be {\mathcal
W}_n^{(d)}(X,\mu,\nu)={e^{iX}\over 2\pi} e^{-{(\mu^2+\nu^2)\over
4}} L_n\left({\mu^2+\nu^2\over 2}\right) \label{dualfockn} \ee
where $L_n$ are Laguerre polynomials; also, the dual symbol of the
transition operator $|n\rangle\langle m|$ has the form \be
{\mathcal W}_{m,n}^{(d)}(X,\mu,\nu)={e^{iX}\over 2\pi}
e^{-{(\mu^2+\nu^2)\over 4}}
 \left({\nu-i\mu\over \sqrt{2}}\right)^{m-n} L_n^{m-n}
 \left({\mu^2+\nu^2\over 2}\right).
\ee For comparison, we report the  symbols related to coherent and
Fock states in the standard tomographic map. They are respectively
 \be {\mathcal
W}_\alpha (X,\mu,\nu)={1\over \sqrt{\pi(\mu^2+\nu^2)}}
 \exp\left[-{(X-\sqrt{2}~\mu~{\rm
Re}~\alpha -\nu~ {\rm Im}~\alpha)^2\over \mu^2+\nu^2} \right] \ee
and
 \be {\mathcal W}_n
(X,\mu,\nu)={1\over \sqrt{\pi(\mu^2+\nu^2)}}\exp\left[-{X^2\over
\mu^2+\nu^2 }\right]
 {1\over 2^n n!}H_n^2\left({X\over \sqrt{\mu^2+\nu^2}}\right)
\ee where $H_n$ are Hermite polynomials.
\section{Conclusions}
To conclude we point out the main results of the work. The duality
symmetry of the star product scheme described by Eqs.
\eqn{deq-map}, \eqn{quant-map}, which provides dual algebras of
operator symbols was elucidated. Some known results which give
raise to special families of symbols relevant for quantum
mechanics, like the quasi-distribution functions of  Husimi
(Q-functions) and Glauber-Sudarshan (P-quasi-distributions) are
shown to be connected via the duality symmetry. The main  result
of the paper is the construction of the dual tomographic map and
the explicit calculation of the kernel for the dual star-product.
This yields  a new noncommutative algebra, the algebra of dual
tomographic symbols.

\end{document}